\def \yskip{\penalty-50\vskip3pt plus 3pt minus 2pt}
\def \reference{\par \yskip \noindent \hangindent .4in \hangafter 1}
\def \abc#1#2#3#4 {\reference#1, {\sl#2}, {\bf#3}, #4}
\def \blank {\lower 5pt\hbox to 0.75in{\hrulefill}}

\def \cm{~\rm{cm}}
\def \s{~\rm{s}}
\def \km{~\rm{km}}

\def \K{~\rm{K}}

\def \AU{~\rm{AU}}

\def \yrs{~\rm{yrs}}
\def \yr{~\rm{yr}}
\def \pc{~\rm{pc}}

\def \erg{~\rm{erg}}
\def \lesssim{\mathrel{<\kern-1.0em\lower0.9ex\hbox{$\sim$}}}
\def \gtrsim{\mathrel{>\kern-1.0em\lower0.9ex\hbox{$\sim$}}}

\documentstyle[11pt,aasms,tighten,flushrt]{article}

\begin{document}

\setcounter{page}{1}

\title {THE ``TWIN JET'' PLANETARY NEBULA M2--9}
\author{Mario Livio\/\altaffilmark{1} and Noam Soker\/\altaffilmark{2} }
\altaffiltext{1}{Space Telescope Science Institute,
3700 San Martin Drive, Baltimore, MD 21218;
mlivio@stsci.edu}
\altaffiltext{2}{Department of Physics, University of Haifa at
Oranim, Oranim, Tivon 36006, Israel; soker@physics.technion.ac.il}

$$
$$

\centerline {\bf ABSTRACT}

We present a model for the structure, temporal behavior, and
evolutionary status of the bipolar nebula M2$-$9.  According to
this model the system consists of an AGB or post-AGB star and a
hot white dwarf companion, with an orbital period of about 120
years. The white dwarf has undergone a symbiotic nova eruption
about 1200 years ago, followed by a supersoft x-ray source phase.
The positional shift of the bright knots in the inner nebular
lobes is explained in terms of a revolving ionizing source. We
show that the interaction between the slow, AGB star's wind, and
a collimated fast wind from the white dwarf clears a path for
the ionizing radiation in one direction, while the radiation is
attenuated in others. This results in the mirror-symmetric (as
opposed to the more common point-symmetric) shift in the knots.
We show that M2$-$9 provides an important evolutionary link
among planetary nebulae with binary central stars, symbiotic
systems, and supersoft x-ray sources.

\bigskip


\section{INTRODUCTION}
M2$-$9 is a spectacular, well studied bipolar nebula, usually
classified as a planetary nebula (PN).  It consists of two oppositely
directed, narrow, cylindrical-like bright lobes, each extending to an
angular projected distance of $\sim 20 ^{\prime \prime}$ from the
center of the nebula (Fig.~1a). Farther out, there are two ``ansae,'' one on
each side along the symmetry axis (Schwarz et~al.\ 1997), at $\sim
60^{\prime \prime}$ from the center, moving outward at $\sim 190 \km
\s^{-1}$ (Schwarz et~al.\ 1997). The inner lobes on the other hand,
expand at a much slower speed, of  $\sim 40 \km \s^{-1}$ (Schwarz
et~al.\ 1997; Solf 2000), or maybe even $15 \km \s^{-1}$ (Trammell,
Goodrich \& Dinerstein 1995; hereafter TGD). The ansae have a point
symmetrical structure (namely, are at symmetrical points when
reflected about the central star).

The nebula and its central region show many interesting properties,
which are summarized in, e.g.\ Balick (1989), TGD, Hora \& Latter (1994),
Phillips \& Cuesta (1999), Solf (2000), and Doyle et~al.\ (2000;
hereafter DBCS). Balick (1989), for example, finds an H$\alpha$
linewidth of $11,000 \km \s^{-1}$, implying the presence of a wind
having a velocity of $\sim 5,500 \km \s^{-1}$ from the nucleus. Balick
further notes that ``the spectrum of M2$-$9's nucleus is more similar to
the slow nova RR~Tel, some symbiotic stars, and Seyfert (type~1.9)
galactic nuclei...'' (see also Solf 2000).

In view of these, and other intriguing properties (see below), it it
important to identify the nature of this nebula and to place it in a
broader evolutionary context.   In the present paper we attempt to do
precisely that. We also address specifically a perplexing positional
(time dependent) shift of the bright knots in the inner lobes, $\sim
8^{\prime\prime}$ from the nucleus (Allen \& Swings 1972), which was
recently studied in detail by DBCS. The shifts in the northern and
southern lobes are in the {\it same\/} direction, i.e.\ there is a
mirror-symmetry rather than a point-symmetry of the bright knots. The
picture that emerges from many studies (e.g.\ Allen \& Swings 1972;
van den Bergh 1974; Goodrich 1991; Kohoutek \& Surdej 1980; and for
more information see \S1 of TGD) is that of two UV illuminating beams
in a mirror symmetry, which move around the central UV source at a
period of $\sim 100 \yrs$.   TGD conducted a detailed study of the
ionization state of the shifting knots. One of their conclusions was
that the off-knots regions are ionized by attenuated UV radiation.
They assumed that the lobes contain rings of higher density, and
concluded that: ``The UV source illuminates a section of the lobe,
causing the rings of material to glow,  producing the knots and
explaining and naturally leading to their ``fixed'' north-south
positions. Obscuring material near the star keeps one side of the
lobe from being illuminated by the full force of UV source.'' TGD
also estimated that the column density of the obscuring material
is $\sim 10^{18-18.4} \cm^{-2}$.

Goodrich (1991; see also Scarrott, Scarrott \& Wolstencroft 1993)
suggested that the radiation that ionizes the knots comes from a
companion, and not from the central star itself. DBCS, on the other
hand, attribute the ionization to a fast jet hitting the bright spots.
As we shall explain in the Discussion (\S4), we identify some
difficulties in the scenario proposed by DBCS.  Instead, we propose
in the present paper a different model, which characterizes M2$-$9 as
a stage in the evolution of interacting, detached binary systems.

The important properties of M2$-$9, as they can be deduced from the
different studies are as follows.
\begin{enumerate}
\item The bolometric luminosity of the central star is not known. The
star is heavily obscured (e.g.\ no point source is observed in the
snapshot HST survey; H.~Bond 2000, private communication). A fit to
the IR spectrum gives a total flux of
$1.1\times10^4$~L$_{\odot}$(D/kpc)$^2$ (S.~Kwok 2000, private
communication), but the distance is rather uncertain
(Schwarz et~al.\ 1997 suggest $\sim650$~pc).
\item Lim \& Kwok (2000) argue that the mass loss rate
from the central source is
$\ll 10^{-5}$~M$_\odot \yr^{-1}$, and maybe even below
$10^{-6}$~M$_\odot \yr^{-1}$.
\item The two ansae at very large distances from the lobes expand at
$\sim 200 \km \s^{-1}$ (Schwarz et~al.\ 1997), a high velocity
compared with that of the lobes which expand at 15--$40 \km \s^{-1}$
(Solf 2000).
\item The ionizing source is very probably a WD companion (e.g.\
Goodrich  1991; Schwarz et~al.\ 1997; Lim \& Kwok 2000).
\item Material moving at $\gtrsim 5,000 \km \s^{-1}$ is found near the
center of the nebula (Balick 1989).  This wind was suggested to result
from an accretion disk around the WD companion (e.g.\ Torres-Peimbert \&
Arrieta 1998). This suggestion is consistent with the general finding
that outflow speeds are always of the order of the Keplerian speed
around the central object (Livio 1999, 2000).
\item The ionizing radiation escapes mainly in two directions, north
and south, with an inclination of $\sim 30^\circ$ to the symmetry axis.
The inclination to the symmetry axis is in the {\it same\/} direction
for both ``beams'' (west for both, or east for both; Fig.~1a).  The beams
rotate
around the central star with a period of $ 120 \yrs$ (DBCS).
\item The ionizing radiation does reach other regions around the lobes,
but it is attenuated in those directions (TGD).
\item The Galactic latitude of M2$-$9 is the largest in the list of 43
bipolar PNe of Corradi \& Schwarz (1995). For an assumed distance of
$650 \pc$ it implies a Galactic height of  $200 \pc$. This may imply
that the progenitor of M2$-$9 was not very massive compared to other
bipolar PNe progenitors (perhaps a late~A or~F rather than a B~star).
\end{enumerate}

We argue that {\it all\/} of these properties can be adequately
explained in the context of a model in which a central binary system
that is a supersoft X-ray source (SXS), is in the transition phase from
a symbiotic nebula (possibly generated by a symbiotic nova; Solf 2000)
to a PN. In particular, we demonstrate that a system consisting of an
asymptotic giant branch (AGB) or a post-AGB star (with its associated
slow wind) and a hot WD (which has a fast wind), in an orbital period
of $\sim120$~yrs, can ionize the ``walls'' of the lobes in one direction,
while the ionization is substantially attenuated in other directions.
The revolving ionizing source is explained by a combination of the
attenuating effects of the post-AGB star's wind and by the effects of a
collimated fast wind (CFW) blown by the WD companion, which is bent by
the slower wind from the post-AGB star. This type of flow, where the CFW
is being strongly pushed upon and bent by the AGB (or post-AGB) wind is
termed a `weak-CFW' flow by Soker \& Rappaport (2000, hereafter SR00).
The bent ``jets'' can both compress the ring in each lobe, and clear a
way for the ionizing radiation in the two specific directions around the
symmetry axis. The circumbinary material has special properties (in
terms of its optical depth, see below) for a relatively short period
of time (of the order of a few hundred years), and this enables the
formation of the circulating ionizing beams.

Our model provides an important link between planetary nebulae and
symbiotic nebulae.  The fact that these two types of nebulae may be related
in their shaping mechanisms (and that the nebulae are sometimes
misclassified) has been known for some time (e.g.\ Morris 1987, 1990;
Corradi
1995; Corradi \& Schwarz 1995; Schwarz \& Corradi 1992; Corradi et~al.\
2000). Furthermore, our proposed model for M2$-$9 emphasizes the important
role played by binaries in the shaping of nebulae in general, and in the
formation of jets in particular. The formation of collimated jets in some
of these systems, as evidenced, for example, by the recent observations of
He2$-$90 (Sahai \& Nyman 2000), provides important clues for the processes
by which astrophysical jets are accelerated and collimated in general
(Livio 2000).

Our dynamical model is presented in \S2, the properties of the ionizing
beams are discussed in \S3, and a summary and discussion follow.

\section{AN INTERACTING WINDS MODEL}

We consider the following configuration for M2$-$9, which qualitatively
describes D-type symbiotic systems. D-type symbiotic systems are wide
binaries (orbital periods $\sim 100 \yrs$) in which a white dwarf
accretes from the wind of an AGB star or a Mira variable.
 We propose that in M2$-$9 a WD companion
accretes mass from the wind of an AGB or post-AGB star. An accretion disk
is formed, and this disk powers a mildly collimated fast wind (CFW), or a
jet (see e.g.\ Livio 1999, 2000 for a review of jet formation). The general
geometry is presented schematically in Figure~1b. For a typical post-AGB
star of 0.6~M$_\odot$ and a WD companion of 0.8~M$_\odot$, and an orbital
period of 120 years (DBCS), the orbital separation is $27 \AU$. We assume
a circular orbit, since an eccentric orbit would have caused the nebula
to possess a strong departure from axisymmetry (Soker, Rappaport \& Kiger
2001).  For a slow wind velocity of $15 \km \s^{-1}$ the Bondi-Hoyle
accretion rate is $\sim 0.01$ of the mass loss rate by the post-AGB star.
The latter is $\sim10^{-6}$~M$_{\odot}$~yr$^{-1}$ in the case of M2$-$9
(Lim \& Kwok 2000), assuming a spherical, slow wind. The condition for
the formation of an accretion disk around the WD can be expressed in the
form $R_d\ga R_{\rm WD}$, where $R_d$ is the outer radius of the disk that
would form given the wind's specific angular momentum. The disk radius is
given approximately by (e.g.\ Livio \& Warner 1984)
\begin{equation}
R_d\simeq10^{10}\eta^2
\left({{M_{\rm tot}}\over {2~{\rm M}_{\odot}}}\right)
\left({{a}\over{10^{15}~{\rm cm}}}\right)^{-3}
\left({{M_{\rm WD}}\over{{\rm M}_{\odot}}}\right)^3
\left({{V_{\rm rel}}\over{20~{\rm km~s}^{-1}}}\right)^{-8}~{\rm cm}~~,
\end{equation}
where $M_{\rm tot}$ is the total system mass, $a$ is the separation,
$V_{\rm rel}$ is the relative velocity between the WD and the wind, and
$\eta$ is a dimensionless parameter determined by the existing density
or velocity gradients in the wind. Typically $\eta\sim0.25$ (e.g.\
Ruffert 1999), which for the assumed parameters gives
$R_d/R_{\rm WD} \sim 25$.
Thus, for the assumed parameters a disk can form, but the
collimation process may not be extremely efficient, since efficient
collimation requires $R_d/R_{\rm WD}\gg1$ (e.g.\ Spruit 1996; Livio
1999). If we assume collimation to take place by a poloidal magnetic
field [of the form $B_z\sim(r^2/R_{\rm WD}^2+1)^{-1/2}$] threading the
disk, then the minimum jet opening angle is given by
$\theta\gtrsim(R_{\rm WD}/R_d)^{1/2}\gtrsim 10^{\rm o}$. This conclusion
is quite consistent with the geometry of the inner lobes in M2$-$9,
which gives an opening angle (or collimation angle, defined from the
symmetry axis) of the CFW of $\theta_c\simeq35^\circ$. We assume that
roughly 1\% of the accreted mass is blown into the CFW (this is
consistent with low-efficiency disk outflows; Livio 1999).
For the CFW velocity we take $v_f = 5,000 \km \s^{-1}$ (of the order of the
Keplerian velocity near the WD surface and consistent with observations;
e.g.\ Livio 2000, Balick 1989).
 We find from these values, and from the assumption that the mass
loss rate from the AGB, or post AGB, star in the equatorial plane
is somewhat higher than that along the polar directions (SR00)
that the ratio of the slow wind momentum flux ($\rho v^2$) to
that of the CFW is $\sim 5$, hence the CFW will be bent and the
slow wind will flow along and near the symmetry axis.
For the assumed parameters, the slow wind impacts the
CFW at a large angle, and the bending is less efficient than in the case
$a \gg h$ (where $h$ is the height of the impact point where substantial
bending occurs; Fig.~1b).  Bending to the asymptotic angle $\theta_d$ will
therefore occur at a height $h=$~a few~$\times h_0\sim 100 \AU$ where
$h_0\sim21$~AU is given by eq.~12 of SR00 (see Fig.~1b).

The CFW passes through a strong shock and forms a hot bubble which
compresses and accelerates the slow wind with which it collides, thus
forming a dense shell (e.g.\ SR00, Fig.~1b).  It is important to note
though that by ``shell'' we do not refer to a closed elliptical or
spherical shell, but rather to a ``corkscrew'' structure, or a ``snail''
type structure. Because of the relatively low mass loss rate from the
post-AGB star, even the dense shells are rather tenuous, and hence, too
faint to be clearly observed. If we assume that all the kinetic energy
of the CFW is transferred to the slow wind material along the lobes,
which for our assumed collimation angle is of the order of $\epsilon
\sim 0.2$ of the slow wind mass, we find the shell velocity to be
\begin{eqnarray}
v_{\rm shell} \simeq
v_s \left[ 1+ {{\dot M_f v_f^2} \over {\epsilon \vert \dot M_s \vert
v_s^2}} \right]^{1/2} \simeq 190
\left( {{\dot M_f} \over {0.001 \epsilon \vert \dot M_s \vert }}
\right)^{1/2}
\left( {{v_f} \over {400 v_s }} \right)
\left( {{ v_s} \over {15 \km \s^{-1}}} \right) \km \s ^{-1}.
\end{eqnarray}

However, the acceleration of a dense shell by a more dilute hot bubble is
prone to Rayleigh-Taylor instabilities, and the shell most likely will
break up into many clumps.  This has three important effects. First, the
acceleration is much less efficient since the hot, shocked CFW leaks
between these clumps. The terminal shell (clump) velocity may thus be as
low as $v_{\rm clump} \sim 50 \km \s^{-1}$. Second, the clumpy shells (a
shell each orbital period) allow more of the ionizing radiation to reach
the lobes. Finally, the leakage of the CFW keeps the directionality of
the expanding bubble to be mainly radial. The clumps reach the bright
rings on the lobes (where the N2 and S2 knots are located; see TGD) on a
timescale of
\begin{eqnarray}
t_c =  500
\left( {{D_r} \over {0.025 \pc}} \right)
\left( {{v_{\rm clump}} \over {50 \km \s^{-1}}} \right)^{-1} \yrs,
\end{eqnarray}
where $D_r$ is the distance to the bright rings in the lobes. The
dynamical age of M2$-$9 is $\sim 1,200 \yrs$ (Schwarz et~al.\ 1997),
hence the first shell reached the ring $\sim 700$~years ago. For an
orbital period of 120 years this means that $\sim 6$ shells have
reached the rings. We propose that the high density material in the
rings was formed by these clumps (see also TGD).

The following thing should be noted in regard to the mass loss from the
WD. As the WD accretes from the wind, it could undergo periodic
symbiotic nova outbursts (e.g.\ Livio, Prialnik \& Regev 1989). Phases
of a strong WD wind could follow such outbursts (e.g.\ Kovetz 1998 and
references therein). In fact, all recorded symbiotic nova eruptions were
followed by an intensive Wolf-Rayet type wind from the hot companion
(Mikolajewska 2000). The recurrence time for such outbursts is given
approximately by (Livio 1994)
\begin{eqnarray}
\tau_{\rm rec} &\simeq &8.7\times10^3
\left( {{\dot{M}_{\rm acc}}\over{10^{-8}~{\rm M}_{\odot}~{\rm
yr}^{-1}}}\right)^{-1}\times\nonumber\\
&&\left[1.54\left({{M_{\rm WD}}\over{{\rm M}_{\odot}}}\right)^{-7/3}
-2.0\left({{M_{\rm WD}}\over{{\rm M}_{\odot}}}\right)^{-1}
+0.65\left({{M_{\rm WD}}\over{{\rm
M}_{\odot}}}\right)^{1/3}\right]^{0.7}
\yrs ~~.
\end{eqnarray}
Thus, for a 0.8~M$_{\odot}$ WD we can expect strong WD wind phases
roughly every $\sim6700$~yrs. We propose that such an outburst formed the
``ansae'' of M2$-$9 $\sim 1200$ years ago. The long time that has elapsed
since the eruption implies, in the case of M2$-$9, a current lower mass
loss rate and a faster velocity than in typical post eruption symbiotics.
We note that the formation of jets or any ballistic outflows, in short
bursts, leads to a linear relation between the distance from the center
and the outflow velocity. Such a linear relation is observed in many
bipolar PNe, suggesting that bursts are a common phenomenon in bipolar PNe.

Another important aspect of symbiotic-nova outbursts is the fact that
following eruptions, the WD enters a supersoft x-ray phase, in which
it is characterized by an effective temperature of 150,000--600,000~K
(e.g.\ DiStefano \& Rappaport 1994; Yungelson et~al.\ 1996). During this
phase, the ionization balance of the circumbinary material is determined
by the radiation from this embedded supersoft x-ray source (SXS). Fast
collimated outflows have been observed in three of the SXSs:
RX~J0513.9$-$6951 ($3800 \km \s^{-1}$, Southwell et~al.\ 1996; Southwell,
Livio, \& Pringle 1997); RX~J0019.8+2156 ($\sim 1000 \km \s^{-1}$, Becker
et~al.\ 1998); and RX~J0925.7$-$4758 ($5200 \km \s^{-1}$, Motch 1998).
The mass transfer rates in these systems are not known precisely, but are
probably above $10^{-8}$~M$_\odot \yr^{-1}$ in all three sources.

In principle, the WD can also be a more permanent SXS, similar in nature
to systems like CAL~83 and CAL~87. In these systems, the accretion rate
onto the WD is sufficiently high for the hydrogen to burn steadily (e.g.\
van den Heuvel et~al.\ 1992). For a WD of 0.6--0.8~M$_{\odot}$, the
accretion rate for steady burning is of the order of (a few)~$\times
10^{-8}$~M$_{\odot}$ yr$^{-1}$ (e.g.\ Nomoto 1982), definitely achievable
by Bondi-Hoyle accretion from the post AGB star's wind. Steady burning
will result in a soft x-ray luminosity of
\begin{equation}
L_x\simeq 10^{37}
\left({{\dot{M}_{\rm AGB}}\over{10^{-6}~{\rm M}_{\odot}~{\rm
yr}^{-1}}}\right)
\left({{M_{\rm WD}}\over{{\rm M}_{\odot}}}\right)^2
\left({{a}\over{27~{\rm AU}}}\right)^{-2}
\left({{V_W}\over{15~{\rm km~s}^{-1}}}\right)^{-4}~{\rm erg~s}^{-1}~~,
\end{equation}
and an effective temperature of
\begin{equation}
T_{\rm SXS}\simeq5\times10^5
\left({{L_x}\over{10^{37}~{\rm ergs~s}^{-1}}}\right)^{1/4}~{\rm K}~~.
\end{equation}

\section{THE IONIZING ``BEAMS''}

In order for the soft x-rays to reach the outer visible parts of the
M2$-$9 nebula (and cause them to fluoresce), they must ionize their own
path through the stellar wind of the AGB star and other intervening
matter, and/or have a path cleared by the action of the CFW coming from
the white dwarf. Given that much of the outer nebula apparently receives
a substantial ionizing flux, we conclude that the radiation is able to
ionize its way through the stellar wind in many if not most directions.
A number of works have addressed the question of the ionization structure
and concomitant line emission from an ionizing source embedded in a
stellar wind (see, e.g.\ Nussbaumer \& Vogel 1987; Seaquist, Taylor \&
Button 1984). Here we are mainly concerned with the fraction of the
radiation flux, as a function of direction, that can pass through the
wind and which is available to illuminate the outer nebula. This
radiation flux can be estimated as follows.

Even though the ionizing radiation is emitted by the WD companion, while
the slow wind is blown by the AGB or post-AGB star, the geometry can still
be safely assumed to be nearly spherical. This is a consequence of the fact
that the CFW clears the slow wind close to the WD, up to a distance of the
order of the bending distance (see previous section), of $r_{\rm min} \sim
100 \AU$, which is considerably larger than the orbital separation of $a
\sim 27 \AU$. For the ionizing radiation to be able to reach the lobes, it
needs to ionize both the newly ejected slow wind material, and the
recombining wind, in its way. For the parameters of the flow discussed here
the newly ejected material is not important. The total recombination rate
from a minimum distance $r_{\rm min}$ to the lobes at $r_l \gg r_{\rm min}$
is given by (assuming a spherical flow)
\begin{eqnarray}
\dot R = 4 \pi \int_{r_{\rm min}}^{r_l} \alpha n_i n_e r^2 dr =
8 \times 10^{45}
\left( {{\dot M_{\rm AGB}} \over {10^{-6}~{\rm M}_\odot \yr^{-1} }}
\right)^2
\left( {{v_s} \over {15 \km \s^{-1} }} \right)^{-2}
\left( {{r_{\rm min}} \over {100 \AU }} \right)^{-1}
\s^{-1}~~,
\end{eqnarray}
where $\alpha$ is the recombination coefficient, $n_e$ and $n_i$ are
the electron and ion  number densities (respectively). Here we assumed
a fully ionized nebula of solar composition, and for the density we used
$\rho=\dot M_{\rm AGB} [4 \pi r^2 v_s]^{-1}$. The number of ionizing
photons emitted by the WD per unit time, assuming a temperature of
$T_{\rm SXS} \simeq 5\times10^5 \K$, is
\begin{eqnarray}
\dot S \simeq 4 \times 10^{46}
\left( {{L_x}\over{10^{37} \erg \s^{-1}}} \right)
\s^{-1}.
\end{eqnarray}

The two rates obtained above, for $\dot R$ and $\dot S$, are within an
order of magnitude from each other.  This implies that a substantial
fraction, but not all, of the radiation will be absorbed along
directions facing the AGB star, while only a negligible fraction will
be absorbed along directions cleared by the CFW. The CFW clears away
the slow wind along its path. If the compressed shell moves at $\sim 50
\km \s^{-1}$ (eq.~2 and following discussion), then during about half an
orbital period (60 years), before fresh slow wind material enters the
swept-up region, it will move to a distance of $r \sim 600 \AU$. This
means that the total number of recombinations along that direction is
much smaller than the number of ionizing photons emitted in that
direction. Along other directions, the slow wind's closest distance to
the WD will be less than the distance of bending, and a substantial
fraction of the ionizing flux will be absorbed, but some will still
reach the lobes. This is consistent with deductions from observations
(TGD). The estimates presented by eqs.~(7) and~(8) therefore support
our proposed scenario, in which the east-west motion of the bright knots
is caused by ionizing beams which pass through channels cleared by bent
``jets'' (or a CFW).

\section{DISCUSSION AND SUMMARY}

M2$-$9 is a unique nebula in that it shows a variability on a time scale
which is consistent with the orbital period expected from the binary
model for the formation of bipolar nebulae. In this model most (although
not all) bipolar nebulae are formed by binary systems with orbital
periods in the range of $\sim 1$--$10^3 \yrs$ (e.g.\ SR00). The time
varying features exhibit a mirror symmetry, rather than a point-symmetry
(TGD; DBCS). As we have shown, a mirror symmetry is expected from the
interaction of two winds, a slow wind emitted by the AGB or post AGB star,
and a jet (or collimated fast wind [CFW]) emanating from the accretion
disk around the WD companion. The mirror symmetry is produced as the CFW
is pushed and bent by the slow wind. Our model identifies the evolutionary
state of M2$-$9 as one which relates three different classes of binary
systems: supersoft X-ray sources (SXSs), symbiotic nebulae, and PNe (or
proto-PNe). The ionizing source is the white dwarf companion, and not the
mass-losing star (as already concluded by e.g.\ Goodrich  1991; Schwarz
et~al.\ 1997; Solf 2000; Lim \& Kwok 2000). At the same time, the fast
outward moving ``ansae'' (Schwarz et~al.\ 1997) and the large and bright
nebula strongly suggest that in the recent past the system has both
undergone a more intense mass loss episode, and an outburst associated
with the accreting WD companion.  Hence, our evolutionary scenario for the
nebula can be characterized as a symbiotic nebula evolving to a PN.

While M2$-$9 is unique in its time variation, it is not in its morphology.
The bipolar proto PN~Henize 401, for example, has a similar structure
(Sahai, Bujarrabal, \& Zijlstra 1999), but with no evidence for circulating
ionizing beams. We note that Sahai et~al.\ (1999) argue that the bipolar
structure in He~401 was formed by collimated bipolar jets, and not by the
more commonly assumed interacting winds model, in which a spherical fast
wind catches up with a previously ejected bipolar slow wind (e.g.\ Balick
1987).  The model of Sahai et~al.\ has in fact some features that are
similar to those of our model for M2$-$9.

Our model offers an interpretation to the bright knots that is different
from the one proposed by DBCS. We showed that the bright knots are exposed
to a higher flux of ionizing radiation (from the hot WD) than the rest of
the lobes, by virtue of the path cleared by the CFW. DBCS, on the other
hand, attribute the brightness/ionization of the knots to shock heating,
caused by jets impacting on lobe walls. We see a few problems with this
suggestion: (i)~it offers no clear explanation for the mirror symmetry
(the jets will collide with slow wind material ejected half a period
earlier and will be slowed down substantially). (ii)~The fact that the
off-knots regions are ionized as well (TGD) argues for an ionizing source
that is not confirmed to the small cross section of a jet, which is present
at all times. (iii)~We find the `corkscrew' model of DBCS (see e.g.\ their
Fig.~2) to provide for a rather poor description of the nebula (e.g.\ the
lines that are supposed to represent the emission in 1989 and 1999 do not
follow very well the actual emission). We rather interpret the 1989 and
1999 images as follows.  The bright emission regions (from the center of
the nebula to the bright blobs along the symmetry axis) are moving in the
same sense, with the 1989 emission being toward the outskirts (east side)
relative to the 1999 emission. The change in the east-west location occurs
only near the bright blobs (N$_3$ and S$_3$), and even there it represents a
complicated flow structure, and radiative transfer from these regions
outward. Nevertheless, our model does share some common features with that
of DBCS (e.g.\ a fast outflow from the companion).

One point in our model deserves further discussion. A cleared path in a
certain direction for the ionizing beam (and therefore enhanced ionizing
radiation), could result, in principle, in a more extended, rather than
brighter, nebula (which is not observed). We should therefore check that
for the parameters of M2$-$9 there is no significant penetration of the
ionization front in the direction {\it perpendicular\/} to the lobe walls.
For an electron density in the lobes of $n_e\sim10^4$~cm$^{-3}$ (DBCS),
an ionizing source of $\dot S\sim4\times10^{46}$~s$^{-1}$ (eq.~8), and a
path clearing for about 10\% of the period, one can easily show that the
penetration of the ionization perpendicular to the wall would be by less
than $\sim0.1D$ (where $D$ is the radial distance to the S$_2$, N$_2$
knots). Thus, while we do not expect any significant perpendicular
penetration, we do expect the bright regions to extend {\it along\/} the
lobes (due to the lower density), as observed (Fig.~1a).
\newpage 

To conclude, the spectacular nebula M2$-$9 provides an opportunity to
identify an evolutionary link among three types of (empirically defined)
binaries: PNe with binary central stars, symbiotic binaries, and supersoft
x-ray sources.


{\bf ACKNOWLEDGMENTS:}
ML acknowledges support by NASA Grant NAG5-6857.
NS acknowledges partial support by a grant from the
Israel Science Foundation, and a grant from the US-Israel
Binational Science Foundation.  We acknowledge extremely
helpful comments and suggestions from Saul Rappaport.



\section*{FIGURE CAPTIONS}

\noindent {\bf Fig.~1a:}
The general structure of M2$-$9.
\bigskip

\noindent{\bf Fig.~1b:}
A schematic drawing of the flow structure (not to scale).
The figure shows the plane perpendicular to the equatorial plane, and
momentarily containing the two stars. The 3D structure is that of a
corkscrew. Drawn on the figure are:
the slow wind from the post-AGB star (stream lines originating
from the post-AGB star); stream lines of the pre-shocked CFW;
the shock front of the CFW; The hot bubble region (again, a corkscrew
structure in 3D); and the dense shell formed from the slow wind material
accelerated by the hot bubble, with its Rayleigh-Taylor instabilities.


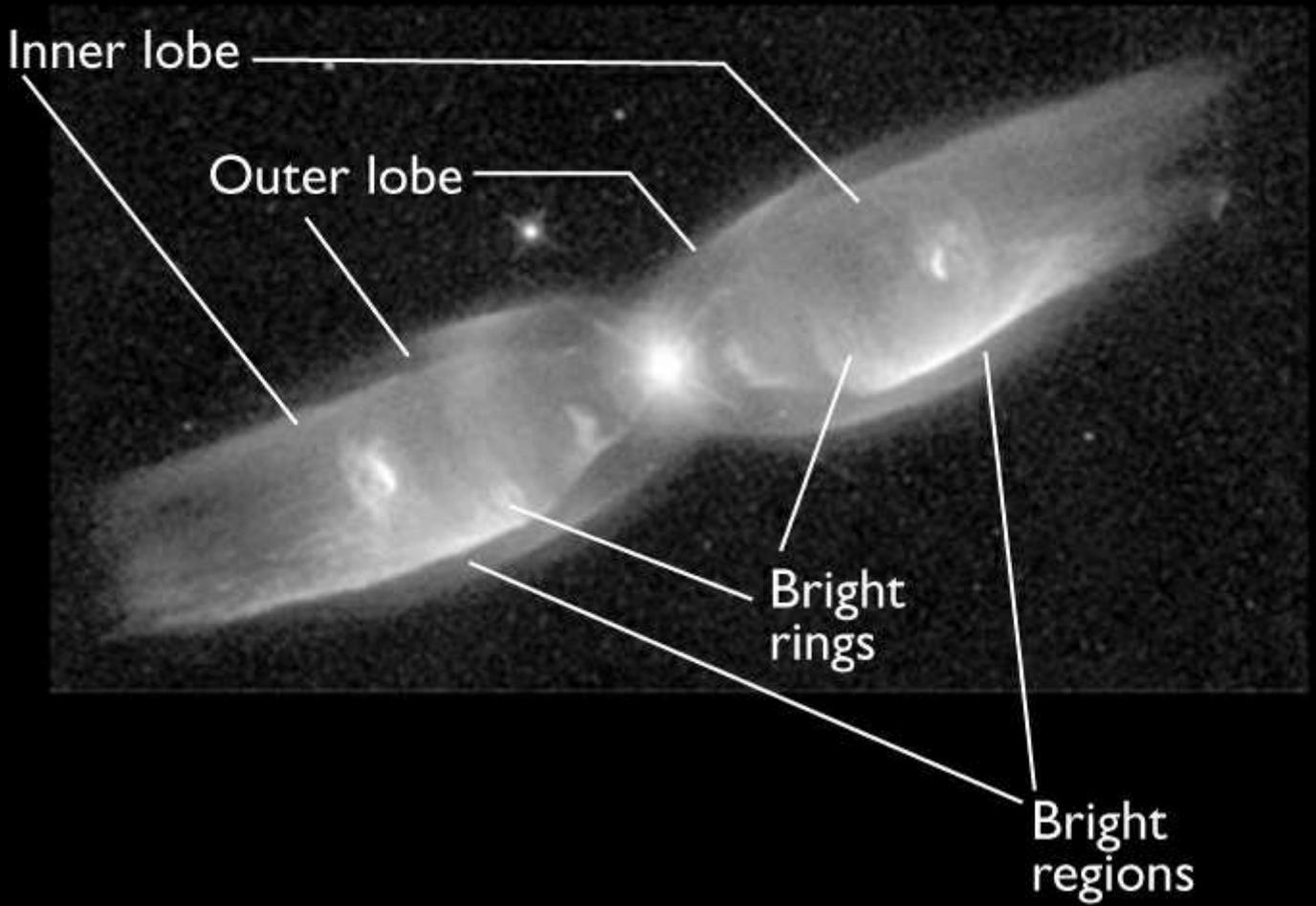
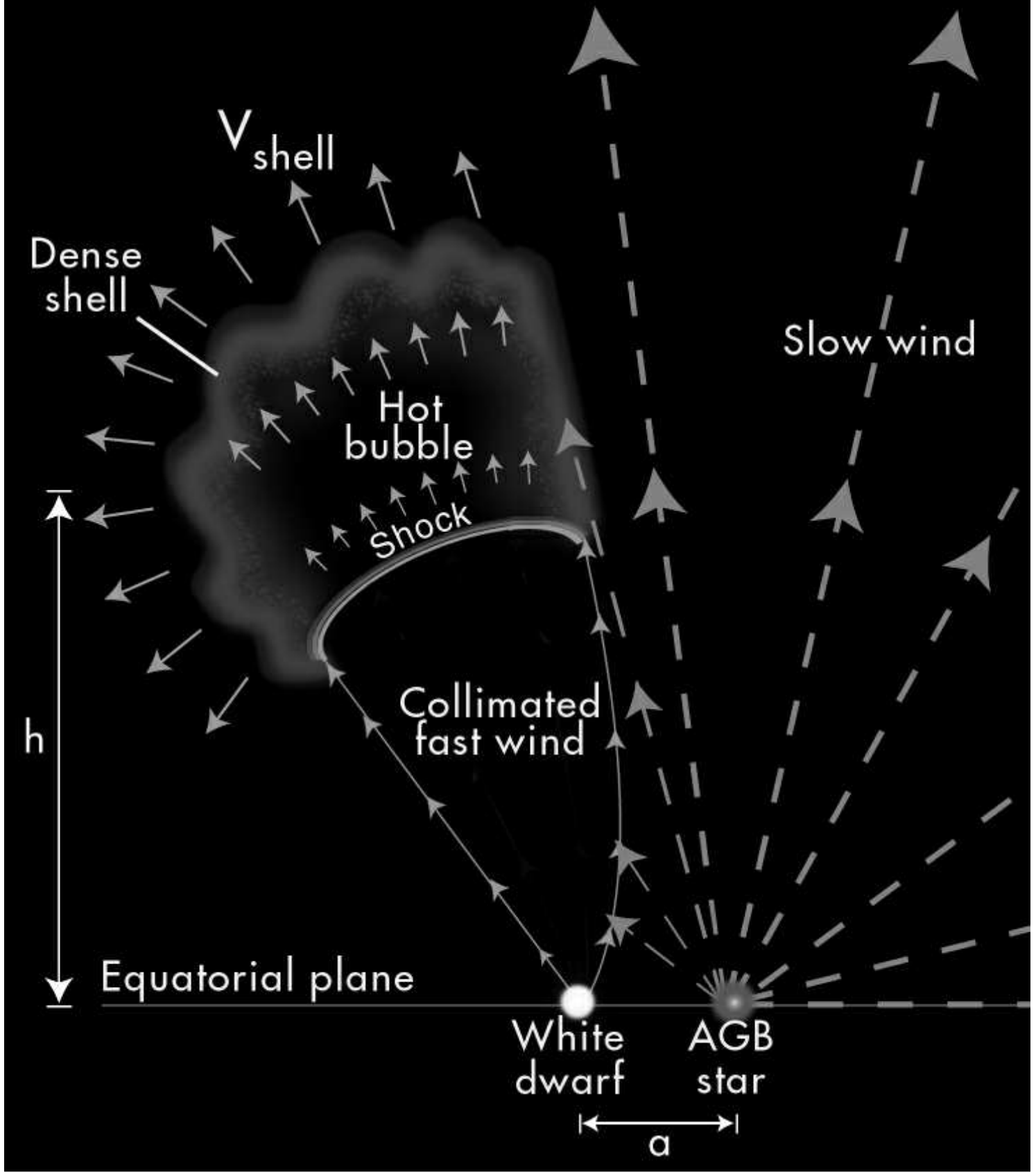
\end{document}